# An IoT Cloud and Big Data Architecture for the Maintenance of Home Appliances

Pedro Chaves, Tiago Fonseca, Luis Lino Ferreira, Bernardo Cabral, Orlando Sousa
*School Of Engineering of the Politechnical Institute of Porto*
Porto, Portugal
{migas, calof, llf, bemac, oms}@isep.ipp.pt

André Oliveira
*Cleanwatts*
Coimbra, Portugal
aoliveira@cleanwatts.energy

Jorge Landeck
*University of Coimbra*
Coimbra, Portugal
jlandeck@cleanwatts.energy

*Abstract*— Billions of interconnected Internet of Things (IoT) sensors and devices collect tremendous amounts of data from real-world scenarios. Big data is generating increasing interest in a wide range of industries. Once data is analyzed through compute-intensive Machine Learning (ML) methods, it can derive critical business value for organizations. Powerful platforms are essential to handle and process such massive collections of information cost-effectively and conveniently. This work introduces a distributed and scalable platform architecture that can be deployed for efficient real-world big data collection and analytics. The proposed system was tested with a case study for Predictive Maintenance of Home Appliances, where current and vibration sensors with high acquisition frequency were connected to washing machines and refrigerators. The introduced platform was used to collect, store, and analyze the data. The experimental results demonstrated that the presented system could be advantageous for tackling real-world IoT scenarios in a cost-effective and local approach.

## I. Introduction

The exponential growth in the number of the Internet of Things (IoT) sensors powered by technological advancements in wireless communication and digital electronics is tightly coupled with the vast amounts of collected data. Reports show that the number of Internet-connected devices is expected to increase from 14.4 billion in 2022 to 27 billion connected IoT devices in 2025 [1], with as many as 73.1 zettabytes of generated data [2], a unit equal to one trillion gigabytes. Big data holds value for organizations, as hidden insights such as trends emerge when analyzing the data, for example, with Machine Learning (ML).

Systems capable of managing big data in an endlessly expanding network face non-trivial concerns regarding computing power, data collection efficiency, real-time data analytics, and privacy. The processing of high volumes of data on IoT applications requires a platform capable of acquiring and storing data from multiple sensors, sometimes thousands. Then, data pre-processing at Edge or Cloud level enables the filtering of outliers and detection of missing data. Such a system also requires the storage of very large amounts of data after each operation over the acquired data.

Only afterward can ML algorithms be applied. From traditional feature extraction ML techniques to the most state-of-the-art deep learning methods used in a variety of areas, the data analysis step is the one that requires the biggest amount of computing power to mine, and extract, useful knowledge from the collected data. Finally, these systems need dashboards to translate extracted insights to users and organizations.

The operations mentioned above are usually done in series, wasting resources and resulting in long delays, making them impractical for time-sensitive applications that need analysis in near real-time. However, multiple computations, e.g., from different sensors, can be parallelized, or even distributed among other machines, to speed up the overall process.

These operations require a lot of computational processing power and large amounts of memory and disk space. However, some AI algorithms can also run much faster on Graphics Processing Units (GPUs) [3]. Just as an example, the data from a washing machine on the work described in [4] requires around 300 MB per washing cycle per machine.

The associated hardware costs are the main problem in treating these data at the edge level. As this kind of processing requires massive amounts of storage and processing power, having a dedicated processing unit for each "thing" or group of "things" that need to be monitored is inefficient. This is especially true in the case of low-cost devices, as spending more resources on data analysis techniques is pointless than on the device itself. Cloud solutions are brought into play because resources can be shared dynamically when processing data streaming from multiple sources.

There are a few popular IoT Cloud providers that offer services to help with the implementation of solutions for data analysis. But again, these solutions are usually very expensive or outright impossible to afford, as the pricing of these services is proportional to the number of messages exchanged, the storage space required, and the kind of tools to be used, among other factors.

The architecture introduced in this paper tries to reduce these costs by creating a framework based on a set of open-source tools. These allow the creation of an Edge or Cloud-based infrastructure on diverse types of hardware and use the available computational power of computers designated for other purposes. Furthermore, our architecture can manage, analyse and transform raw data into valuable insights. The solution is also scalable as it can split the workload throughout multiple machines in a way that considers the load state of the available machines over which the computation can be split. Similarly, all machines are interconnected by a communications middleware, responsible for distributing the work, providing an extra layer of reliability to the system.

For the evaluation of the system, a real-world scenario was used, which is based on a predictive maintenance system for home appliances introduced in [4].

The present paper is structured as follows. Section II provides a survey on the state of the art of IoT Cloud architectures for big data processing. Section III describes the main concepts of the architecture being proposed, which is used on a home appliances predictive maintenance application described in Section IV. Finally, Section V addresses the main conclusions and future work.

## II. STATE OF THE ART

The Cloud paradigm offers a dynamic and scalable resource-sharing platform that provides IoT services. Several cloud providers offer solutions for data collection, processing, storing, and visualization. For example, Amazon Web Services (AWS) delivers the IoT Core service, which can connect and exchange messages with sensors via MQTT and HTTPS, among other protocols. AWS IoT Core can then be connected with other services providing storage, computing, and machine learning capabilities, such as Amazon S3, Lambda Functions, and Sagemaker services.

Similarly, Microsoft Azure IoT cloud offers industry-specific services such as Microsoft Cloud for Manufacturing focused on device management, with particular attention to industrial hardware and devices. Devices are connected through the Azure IoT Hub, data is stored with the Azure Data Lake, computing is supported by Azure functions, and machine learning capabilities can be added with Azure ML [5] [6].

Google Cloud is primarily focused on data management/enablement services. Google Cloud IoT Core is similar to Microsoft Azure's IoT Hub. However, it integrates Google's generic analytics tools to enable more data processing and management capabilities [5]. Several other IoT cloud providers exist, such as IBM Watson IoT Platform, Oracle IoT, Siemens MindSphere, BOSCH IoT Suite, and Cisco IoT Cloud Connect [7].

These three major platforms offer simulators to estimate the cost of using their services. For this estimate, 500 washing machines were considered, which would generate roughly 40 thousand million messages per month, and after processing, the data stored would increase by around 300GB each month. To connect, collect, process, and store this data, it was estimated based on the simulations that the cost per month per connected device would be 1€ in Azure [8], 2€ in Google Cloud [9], and 30€ in AWS [10]. AWS pricing greatly increases with high message count values, collecting data at 1KHz for 500 devices with modest utilization, generates the mentioned 40 thousand million monthly messages, and the pricing scheme does not expect this kind of utilization. These values are only a rough estimate but provide a comparison baseline between the platforms and some awareness of the costs of relying on third-party cloud providers.

However, while all these cloud providers supply advanced IoT cloud services, they are also expensive and therefore not suitable for all use case applications. For example, the service budget might be calculated depending on messages between devices and the server, while additional data storage and processing costs may apply.

Several works did also propose some solutions to this kind of problem. For example, in [11] the authors propose a software architecture for processing big-data analytics considering the complete compute continuum, from the edge to the cloud, supported by an elasticity concept that enables these systems to satisfy the performance requirements of extreme-scale analytics workloads, In addition, those solutions must be able to meet non-functional requirements such as real-time, energy-efficiency, communication quality, and security.

Also, in [12], the authors support their infrastructure on the SCoT's architecture, which is an IoT platform designed to support dynamic and large-scale IoT scenarios, but its performance and complexity may compromise the platform. Other solutions, like Arrowhead [13], are particularly interesting in support of interoperability between different protocols and in developing complex service-based applications but lack the capabilities to support elastic cloud applications.

An extensive survey of existing Cloud-based IoT platforms is made in [6], and some of the problems of these platforms are highlighted. As such, an alternative architecture that aims to be generic was also developed by the authors.

With the attempt to broaden the usage of IoT technologies to less costly areas, the maintenance and development costs in most commercial platforms are uneconomical. Specifically, in our case study presented in Section IV, the cost of testing and processing data collected at a data rate of 2048 samples per second would be very expensive. As such, in this paper, we present a scalable IoT platform capable of addressing IoT connections and big data processing in different types of hardware.

## III. ARCHITECTURE

One of the main requirements for the architecture is to be able to grow according to the needs of the application, by adding more processing cores or by adding more computers linked by a high-speed network. The system should also be able to dynamically distribute the load among cores and computers to maximize the system throughput. Finally, it should also offer a flexible, simple, and low-cost solution based on open-source software.

According to the literature and our own experience, we assume that the applications running in such a system will primarily be structured as a pipeline of sequential processes. Nevertheless, simple adaptations to the solution being implemented can also support more complex and diverse structures.

To fulfill these requirements, the envisaged architecture connects the different modules through a communications middleware due to its flexibility and inherent scalability of using the producer/consumer communication model and message queues. This allows each process on the pipeline to retrieve messages from the queues and work with what was previously the output of the last process on the queue. This kind of operation also allows the existence of multiple processes in each stage of the pipeline.

### A. Overview

The proposed elastic cloud, depicted in Figure 1, comprises three levels. At the lowest level, we have the database and the file systems.

A middleware connects and orchestrates all communications between the stages of the pipeline through a set of queues $Q_i$. It can also support connection with the database and with the file system.

At the highest level, we have a set of applications $P_i$, which are constituted by different worker modules organized in a pipeline structure, and if required, parallelized, amongst different servers or machines $M_i$.

There are also a set of monitoring probes that gather performance information from every machine. The data collected by these probes is used to perform the operations that grant the elasticity capabilities to the system.

### B. Data Storage

The Database (DB) has multiple purposes since it allows not only the storage of high volumes of input data, but also processed data, the results of the analysis, and some other configuration data. This database should also be capable of processing many transactions per second. In the use case described in Section IV, the data acquisition frequency can reach 2kHz per appliance being monitored (but of course, multiple samples can be joined into a single message, significantly reducing the data rate). If multiple appliances are being monitored, then the number of transactions can be extremely high.

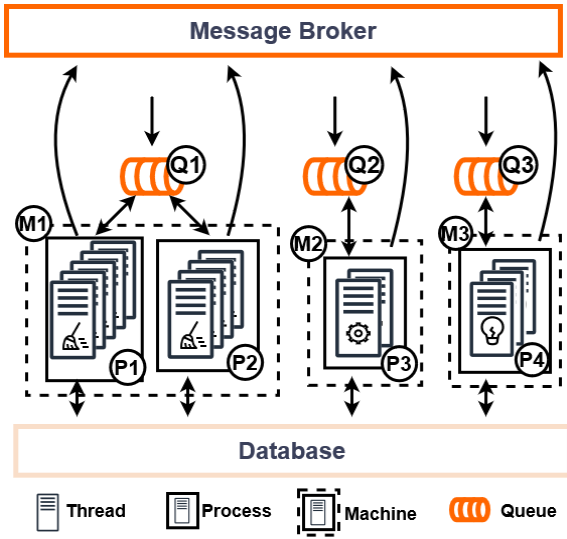

*Figure 1- Architecture overview*

To store the data, various approaches can be followed. Ideally, for time series, a Time Series Database (TSDB) could be used. But a non-relational database combined with a distributed file system is also a very efficient solution, as we discovered during the implementation of the application described in Section V. In such scenarios, data is stored in files (instead of the database), allowing for much higher performance. Furthermore, these files frequently use a comma-separated-values (CSV) format, which is easily accepted by multiple Machine Learning Libraries.

Although the worker modules could access the database through the middleware, we concluded that it is more efficient to directly access the database (placing the required queries), or the file system using a protocol like the File Transfer Protocol (FTP), or a using any other kind of distributed file system.

### C. Middleware and Communication Protocols

Systems like the one being proposed always must support a multitude of different communication protocols particularly to communicate with sensors and with the user interfaces. The Message Broker is the backbone of the system, notifying the different workers when new data is available for processing and when the next step in the pipeline is ready to start. This behavior is supported by the Publisher/Subscriber communication pattern where workers on different processes/machines are linked by a single queue (Figure 1).

This setup allows for the different workers to read data from the same queue, and inherently this allows to distribute the work among multiple workers and on the same stage, either on the same or on different machines. It is also agnostic to the kind of workers being used, enabling the use of Docker containers or similar technologies.

The kind of properties required for the middleware are totally supported by a few middleware technologies, although due to its flexibility, performance, and cost (open source), RabbitMQ has been chosen.

For data acquisition, MQTT is nowadays the most used protocol to communicate with resource-constrained sensors and actuators. Its key features are being lightweight and bandwidth-efficient, making it ideal for restricted environments. Furthermore, it is a messaging protocol that can be framed in the client/server and publish/subscribe patterns, it offers three different levels of quality of service (QoS), where increasing QoS levels improves delivery guarantees at the cost of lower performance and increased use of bandwidth [14]. There are still many sensors that offer connections based on REST APIs. Consequently, this is also a protocol that must be supported for data acquisition, although it should preferably be used for low data throughput applications. The Advanced Message Queueing Protocol (AMQP) allows peer-to-peer communication between processes. Additionally, it defines message formats and encoding [15].

To visualize the data and results obtained by the analysis, a REST API over HTTP is most of the time the ideal solution due to the ease of application development, its management, and the existence of many end-user applications that consume data in this format.

Figure 2 shows one typical solution for communications within the proposed architecture.

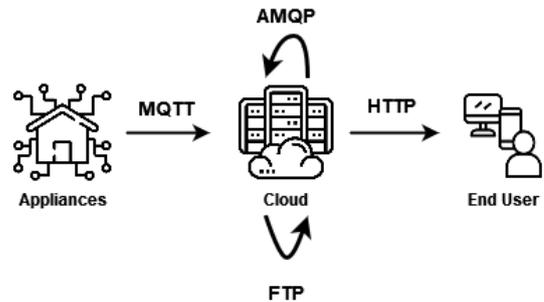

*Figure 2 – Communication schema*

### D. Data Processing Workers

The workers are the entities responsible for performing the tasks at every stage of the pipeline, from data acquisition until the user interface support. Consequently, they will have very different execution characteristics. As an example, workers responsible for data acquisition should have higher communications bandwidth to receive data from multiple sources. Still, these kinds of tasks are usually characterized as being I/O bound, with a reduced processor utilization, since most of the time, they will be waiting for messages. Consequently, it is possible to support multiple workers performing this kind of tasks on the same server. On the other hand, workers who run machine learning algorithms should have access to powerful graphical processing units (GPUs), which enable them to speed up their tasks, and will mostly be characterized by being processor bound.

The instantiation of worker modules on each stage of the pipeline is a process that depends on characteristics of the application being executed and the available resources. The activation, deactivation, and migration of workers to other machines is controlled by the monitoring system as described in Section III.E.

The workers may have different types, but it is expected that all of them communicate through the middleware, they may include: (i) native Linux/Windows applications, (ii) threads or processes from Linux/Windows applications, (iii) containerized applications, running on Docker containers and potentially orchestrated with Kubernets or any other container-based technology, or (iv) cloud services offered by any other cloud platform which is also able to connect to the middleware.

### E. System Monitoring and Elasticity Control

The cloud consists of multiple machines that can be either be active or inactive at a given time, each of these machines can run multiple workers, belonging to the same or to different pipeline stages, according to the application.

Elasticity control can work on a set of elementary principles, like measuring, on each server, its total utilization, and the utilization by each worker and taking decisions accordingly. But other principles could be used, like measuring the execution time of each stage, which would allow to give soft real-time guarantees to the systems.

For the moment, only utilization-based politics are being implemented.

The utilization values are periodically transmitted to a Central Monitoring module which is responsible for the activation of new workers and deactivation of others when not needed. The decisions to activate new workers is based on detecting if the utilization levels are above a specific threshold, for each worker and for each stage on the pipeline. Additionally, the Central Monitor is also able to query the middleware in order to determine the number of elements in each queue, i.e. in order to determine the amount of work in backlog. It is based on this information that new workers are activated on the underutilized servers for the pipeline stages, where they are mainly required.

Multi-threaded and distributed environments make it much harder to detect and manage errors, since it can be hard to identify its source. Therefore, the Local Monitoring agents also play an important role in the detection and management of errors. There are critical points of failure that are important to identify and analyze. The errors are registered, and serious situations should throw notifications with the information needed to rectify the situation. Each server is monitored in real-time, and when an error occurs, the details are written to a log file.

## IV. CASE-STUDY SMART-PDM

The case study focuses on predicting abnormalities in the operation of home appliances through a data-driven approach. If enough data is acquired to train the models successfully for each home appliance model, it is expected that small variations can be detected and associated with a failure or degradation of a certain component. Detecting these failures can minimize costs in the maintenance of the monitored equipment, and in some cases, bigger problems may be prevented. As an example, if the degradation of a bearing is detected on time and action is taken, this might prevent a costly repair or even the retirement of the appliance.

This section details how the introduced big data IoT architecture is applied to this case study, using the distinct parts of the system to store and process large amounts of sensor data in a scalable and distributed manner.

### A. Sensor Data Collection

Home appliances data is collected by the devices introduced in [4], which are sensors capable of acquiring power, temperature, and vibration readings at both slow and high sampling frequencies (configurable from 128Hz up to 16384Hz). In this specific case study, there are two types of time series acquired that we call streams. These types are the slow streams and the fast streams.

The slow streams take one sample per second, while the fast streams collect 2048 samples per second. I.e., a washing machine program run, which takes around 2 hours, will generate approximately 600 MB of data.

The sensor acts as an MQTT (Message Queuing Telemetry Transport) publisher that connects to the MQTT broker. The subscriber receives and stores the readings. As these sensors have a proprietary infrastructure, data is initially stored in an external proprietary database and then fetched to our system through an HTTP API. This process is illustrated in Figure 3.

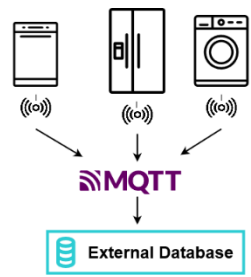

*Figure 3 - Data Collection via MQTT*

### B. Data Processing

The data processing was implemented with the previously introduced architecture, and as such, four different scalable modules are implemented from open-source technologies. RabbitMQ is used as the message broker to connect modules and orchestrate the distributed nature of the implementation. Each worker acts as both a subscriber to the previous worker in the pipeline, and as a publisher to the next one. Each stage has its own queue that can be shared across different machines in a distributed system. As such, the data processing steps can be scaled according to the needs, parallelizing the download of raw data, the data cleaning, and the feature extraction across multiple threads and computers. At the end of each stage in the pipeline, the worker stores its output in a distributed file system, due to the ease of implementation and its performance.

Figure 4 illustrates the case-study data processing steps.

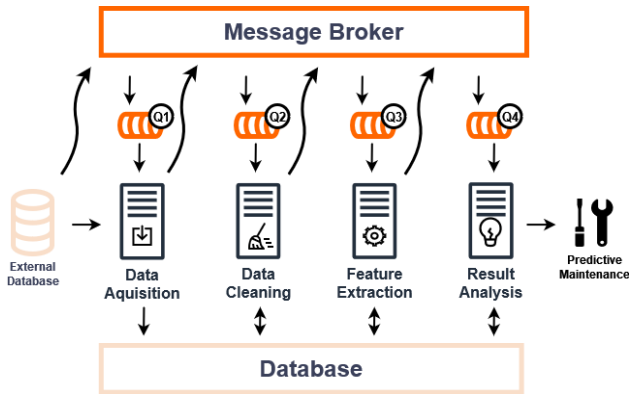

*Figure 4 – Case-study data processing pipeline*

First, the Download Manager worker receives a notification from the external database when there is new meaningful raw data available (e.g., a new washing machine or refrigerator cycle has finished). This notification informs our system about the device identification, the beginning and end timestamps of the cycle. Then the Download Manager proceeds to download the slow stream and the corresponding fast stream, storing both in different CSV files in the File System (FS). Finally, at the end of each downloaded cycle, a notification is sent to the Data Cleaning Queue – queue Q2.

The Data Cleaning worker verifies the data and prepares it to be analyzed. In this worker module, filters to remove noise and outliers can be applied. Other operations might include sorting the data according to its timestamp and removing duplicate readings. This module is also used to detect holes in the data, which are usually due to a temporary lack of connection with the sensors. Nevertheless, in this case, we conclude that even with some holes in the data, the results were very acceptable. After finishing this stage, the Feature Engineering module is notified.

The Feature Engineering module extracts relevant information from the cycles' data, creating features. Extracted features range from very simple ones, such as the maximum, average, and minimum current and vibration to more complex ones, like Fast Fourier Transforms (FFT), Skewness, and Kurtosis. Next, the features are concatenated, grouping the features extracted for the current slow stream with the ones extracted for the current and vibration fast streams into a single feature array that translates the appliance cycle. In this case study, a total of 79 features are extracted for each cycle. In the end, only the most relevant ones are selected and delivered into the result analysis module.

Finally, selected features are used to train machine learning classification models in the Results Analysis module. Once these models attain significant accuracy, they can be deployed to the production environment. In this environment, the same architecture modules are used, as the raw data is collected first, then data cleaning and feature engineering are applied. In the end the trained classification models are applied to classify if that device needs maintenance and what type of failure is happening.

Figure 5 details the described case-study data flow through the implemented system.

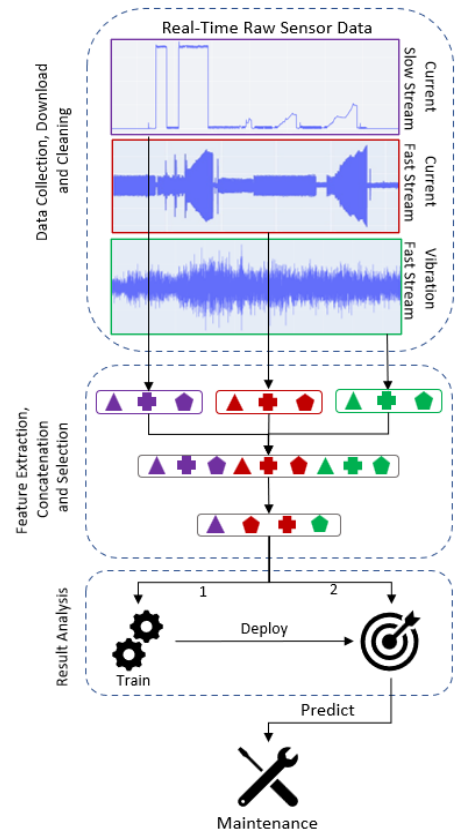

*Figure 5 – Analysis process*

### C. Preliminary Results

Preliminary results using this architecture show a successful implementation of a big data system for home appliances' predictive maintenance. Several appliances, such as washing machines and fridges are currently being monitored in real-time, and their working cycles downloaded, cleaned, and analyzed by the system.

Features extracted from several washing machine cycles, some executed with the machine under normal operation conditions, others with the machine with some kind of malfunction, such as heating and bearings problems, serve as a base to train and compare different types of classification models.

In the first phase of tests, three different ML approaches were trained and optimized to detect the washing machine malfunctions, namely: Support Vector Machine (SVM), Decision Tree (DT), and Random Forest (RF).

All three approaches show similar accuracy results, with the SVM attaining 89.42%, the DT 92.24%, and RF the best overall accuracy of 92.85%. Future work will make use of the implemented architecture to efficiently collect and process more sensor data, improving the training of the classification algorithms.

### D. Dashboard User Interface

Visualizing big data can be a complex task. A file can easily reach millions of lines, each representing a different reading. It is not feasible to extract any meaning from these files in a raw format, so to make it human-readable there needs to be a way of plotting these files. This is particularly useful to check if the system is working properly and identifying periods of data loss or corruption.

In this case-study, a visualization application was developed. It is connected to the *External Database* that performs the data acquisition. Aside from allowing to see the data plotted, it is possible to download the files and to plot local files previously downloaded (Figure 6). Moreover, the dashboard can also show the malfunction predictions by each machine working cycle.

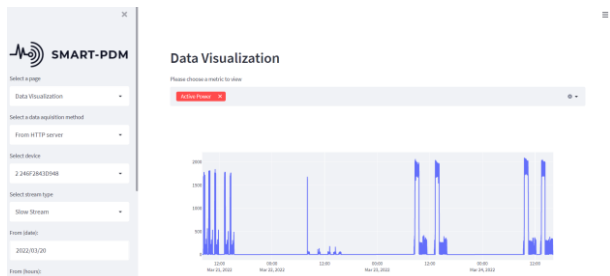

*Figure 6 – User interface example*

## V. Conclusion

This paper proposes an open-source and scalable platform architecture that can be deployed for efficient real-world big data collection and analytics, particularly for IoT applications. It assumes a pipeline architecture for the application, where data is analyzed and transformed in several stages by several distributed or local workers.

The proposed system was tested with a case-study for the Predictive Maintenance of Home Appliances, where current and vibration sensors with high acquisition frequency were connected to multiple home appliances, in this case washing machines and refrigerators. The introduced platform was used to collect, store, and analyse the data. The experimental results demonstrated that the presented system could be advantageous for tackling real-world big-data IoT scenarios in a cost-effective approach. Implementing this system would require an upfront investment as opposed to the monthly fees charged by cloud providers.

This architecture will still be further improved in order to better support the use of Docker components, implement and test more advanced scalability algorithms and adapt the architecture to other kinds of application scenarios.

### Acknowledgment

This work was supported by project SMART-PDM, nº 40123 (AAC nº 25/SI/2017) POCI-01-0247-FEDER-040123, co-funded by the European Regional Development Fund (ERDF), through the Operational Programme for Competitiveness and Internationalization (COMPETE 2020) and also by of project "FERROVIA 4.0", POCI-01-0247-FEDER- 046111, co-funded by the European Regional Development Fund (ERDF), through the Operational Programme for Competitiveness and Internationalization (COMPETE 2020).